\documentclass{ws-p8-50x6-00}

\begin{document}

\title{Multiplicity Fluctuations in DIS}

\author{Leszek Zawiejski \\
for the ZEUS Collaboration }

\address{HEP XII Department, Institute of Nuclear Physics,   
PL-30-055 Krak\'ow, Poland \\ 
E-mail: leszek@chall.ifj.edu.pl}


\maketitle

\abstracts{
The charged multiplicity fluctuations in deep-inelastic scattering (DIS)
are investigated in order to test perturbative QCD
and local parton hadron duality (LPHD). The fluctuations were
measured with the ZEUS detector at HERA 
in restricted phase space domains
in the current Breit frame region.
The measurements are compared to analytic pQCD + LPHD predictions
and QCD-based Monte Carlo models.} 

\vspace*{-0.9cm}
\section{Introduction}
The studies of   
multiplicity fluctuations in restricted domains
of the phase space reveal the nature of particle 
correlations and are sensitive to the dynamics of the underlying process.  
The hadronic final state in DIS is a result of a hard
partonic scattering at a large 
momentum transfers $Q \gg \Lambda$. A subsequent parton cascade
development down to some QCD cut-off $Q_{0}$ 
and fragmentation occuring at small momentum transfers
lead to the observed hadrons. The characteristic QCD scale $\Lambda$ 
is of the order of a few hundred MeV. At present there are two main approaches
for the description of the hadronic final state: the QCD-based Monte Carlo
programs and analytic perturbative QCD calculations \cite{och} in conjunction
with the Local Parton-Hadron Duality (LPHD) hypothesis \cite{ph1}. 
In Monte Carlo programs,
the partonic final state is generated according to the pQCD picture
and below $Q_{0} \simeq $ 1 GeV the transition from partons
to hadrons is performed by applying non-perturbative models. 
In analytic calculations, the parton cascade develops down  
to $Q_{0}$, close to $\Lambda $. In this case the direct comparison of 
the spectra of partons with those of hadrons becomes possible,
without involving any hadronization model.

In the multiplicity fluctuations studies, the method of normalized
factorial moments, $F_{q}$, was used \cite{fm1}.
Experimentally, the moments of order $q$ were calculated by counting
$n$, the number of charged particles in a restricted region    
of phase space, $\Omega$:
\vspace*{-0.1cm}
\begin{equation}
F_q(\Omega) \; = \; <n>^{-q}\; <n(n-1)\ldots (n-q+1)>,
\label{eq:mf1}
\end{equation}
where $< \ldots >$ denotes averaging over all events in the sample.
The phase space region, of size $\Omega$, was defined
either as a polar-angle ring around the jet axis
or as an upper limit on the particle transverse momentum calculated 
with respect to the jet axis.
\vspace*{-0.2cm}
\section{Angular multiplicity fluctuations}
The factorial
moments were measured as a function of the ring width $\theta$ 
in the corresponding cone with half opening angle $\Theta_{0}$ \cite{zeu}.
\begin{figure}[h]
\begin{center}
\epsfxsize=29pc 
\epsfysize=20pc
\epsfbox{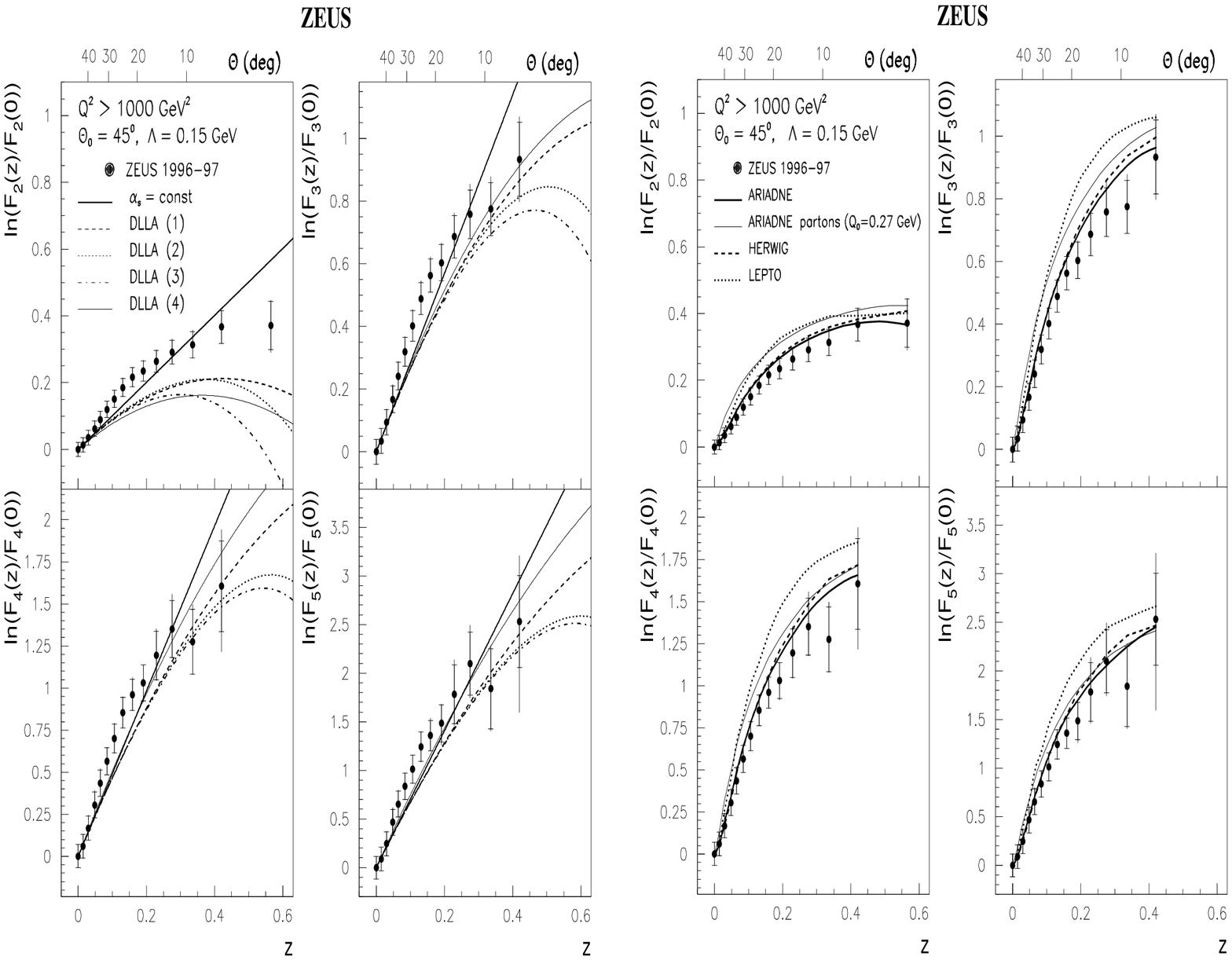} 
\vspace*{-0.8cm}
\caption{(Left): Comparison of the factorial moments with QCD+LPHD calculations.
The curve DLLA (4) includes
a correction from the MLLA in the calculation of $D_{q}$;
(Right): Comparison of the factorial moments
with different MC models at the hadron level.
The parton-level from the ARIADNE MC generator, which is
consistent with LPHD and represents
the analytic calculation, is also shown.}
\vspace*{-0.2cm}
\end{center}
\end{figure}
Substituting $\theta$ with a scaling variable, 
 $z=\ln(\Theta_{0}/ \theta) / \ln(E \Theta_{0}/ \Lambda)$, the 
experimental results can be compared with analytic calculations where 
$\Theta_{0}$ is the half opening angle of a cone around an outgoing quark,
of energy E, radiating the gluons.  
The QCD + LPHD prediction for $F_{q}$ is:
\begin{equation}
\ln \frac{F_q(z)}{F_q(0)}=z\;(1-D_q)(q-1)\;\ln(E \Theta_{0}/ \Lambda),
\label{eq:mf2}
\end{equation}
where $D_{q}$ are $R\acute{e}nyi$ dimensions \cite{red}  
and can be calculated either
in a fixed-coupling regime or in a running-coupling
regime (see \cite{och} and refs. therein) of the Double Leading Log Approximation (DLLA).
For independent particle production:
$D_{q} = 1 $ and $F_{q}(z) = F_{q}(0)$.
Figure 1(left) compares the factorial moments for the DIS
data with the QCD predictions.
A significant disagreement with the data was found. For the higher order moments, 
an improvement is observed particularly for calculations in 
Modified Leading Log Approximation (MLLA) \cite{dq1} (see Fig. 1(left)). 
The data are compared with Monte Carlo models \cite{mc1,mc2,mc3} at the hadron
level in Fig. 1(right).  
All models reproduce the trends seen in the data. The same figure also shows 
the parton level of ARIADNE.
For consistency with the LPHD picture,
the parton cascade was cut-off at $Q_{0}$ = 0.27 GeV, which is close 
to $\Lambda = 0.22 $ GeV.
For higher order moments, the parton level is closer to the data
than the analytic calculations.
\section{Multiplicity fluctuations in limited momentum space}
%
%
According to QCD predictions, soft gluons
in a limited $p_{t}$ region
are independently emitted due to a coherence effect \cite{pt1}.
For the measurements of the $p_{t}^{cut}$ moments, 
Eq. (\ref{eq:mf1}) was used with the requirement that all counted
particles should have transverse momentum $p_{t} < p_{t}^{cut}$.
The $p_{t}^{cut}$ multiplicity moments 
have been calculated using DLLA+LPHD \cite{pt1}. 
For small values of $p_{t}^{cut}$, theory predicts:
\vspace*{-0.1cm} 
\begin{equation}
F_q(p_t^{cut})  \simeq 
1 + \frac{q(q-1)}{6}\>  \frac{\ln(p_t^{cut}/Q_0)}{\ln(E/Q_0)}.
\label{eq:mf3}
\end{equation} 
If $p_{t}^{cut} \rightarrow Q_{0}$, then all $F_{q} \rightarrow 1 $  and the 
multiplicity distribution approaches a Poissonian distribution.  
Figure 2 shows $p_{t}^{cut}$ moments in DIS \cite{zeu} together   
with Monte Carlo predictions 
at the hadron and parton levels. The partons of the ARIADNE MC generator
are consistent with 
analytic calculations. For $p_{t}^{cut}$ $ < 1$ GeV, all moments rise 
in contradiction to the analytic QCD predictions.
The MC results for hadrons show similar trends to the data. 
\begin{figure}[t]
\begin{center}
\epsfxsize=22pc 
\epsfysize 22pc
\epsfbox{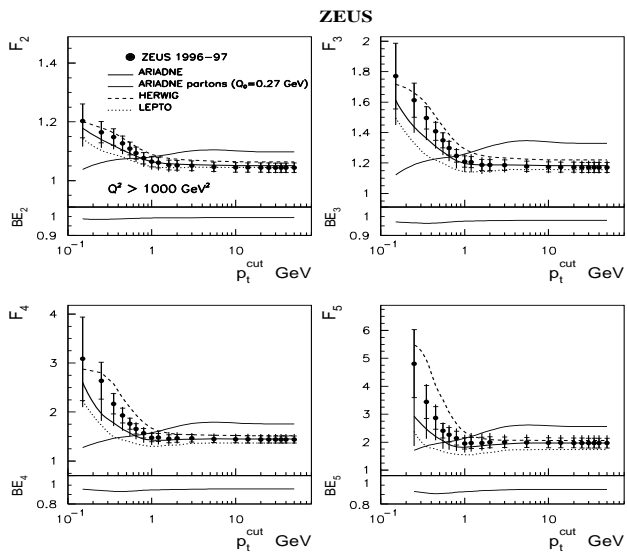} 
\vspace*{-2.cm}
\caption{Comparison of the factorial moments calculated as a function of $p_{t}^{cut}$
with different MC models
at hadron level and parton-level ARIADNE, which represents analytic
calculations and is consistent with LPHD. }
\end{center}
\end{figure}
\vspace*{-0.2cm}
\section{Conclusions}
The multiplicity fluctuations have been measured in DIS in restricted phase 
space regions to test pertubative QCD.
For angular factorial moments, the differences between data and analytic
calculations are quantitative and become smaller for higher order moments
and MLLA corrections. A qualitative disagreement between the QCD+LPHD 
and the data has been observed for $p_{t}^{cut} $ moments. This, for the first time,
indicates a limit of the LPHD hypothesis applied to the multiplicity fluctuations.
The MC programs show  
the large impact of the hadronisation stage on the multiplicity
distributions.
\vspace*{-0.3cm}
\section*{Acknowledgments}
\vspace*{-0.2cm}
I would like to thank the DESY Directorate for support to the
DIS2001 Bologna workshop. Many thanks also to my collaborators at ZEUS especially
S. Chekanov, M. Adamus, B. Foster, J. Whitmore, R. Yoshida and J. Chwastowski for a
careful reading of the manuscript and useful remarks.


\vspace*{-0.4cm}
\begin{thebibliography}{99}
\bibitem{och} V.A. Khoze and W.Ochs, {\em Int. J. Mod. Phys.} A {\bf 12}, 2949 (1997).
\bibitem{ph1} Ya.I. Azimov {\em et al.}, {\em Z. Phys.} C {\bf 27}, 65 (1985).
\bibitem{fm1} E.A. De Wolf {\em et al.}, {\em Phys. Reports} {\bf 270}, 1 (1996).
\bibitem{zeu} ZEUS Collab., S. Chekanov {\em et al.}, {hep-ex/0104036}.  
\bibitem{red} A. R$\acute{e}$nyi, {\em Probability Theory}, North-Holland, Amsterdam, (1970).
\bibitem{dq1} Yu. Dokshitzer and I.M. Dremin {\em Nucl. Phys.} B {\bf 402}, 139 (1993).
\bibitem{mc1} LEPTO, G. Ingelman {\em et al.}, {\em Comp. Phys. Comm.} {\bf 101}, 108 (1997).
\bibitem{mc2} HERWIG, G. Marchesini {\em et al.}, {\em Comp. Phys. Comm.} {\bf 67}, 465 (1992). 
\bibitem{mc3} ARIADNE, L. L\"onnblad, {\em Comp. Phys. Comm.} {\bf 71}, 15 (1992).
\bibitem{pt1} S. Lupia, W. Ochs and J. Wosiek, {\em Nucl. Phys.} B {\bf 540}, 405 (1999).
\end{thebibliography}
\end{document}